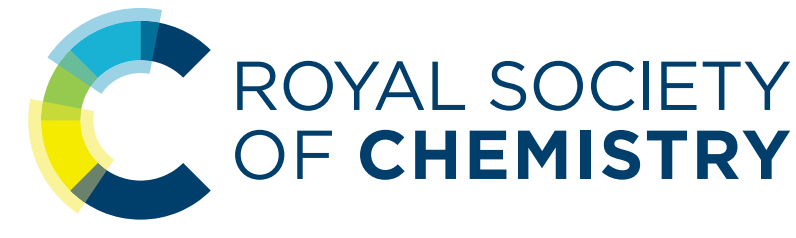

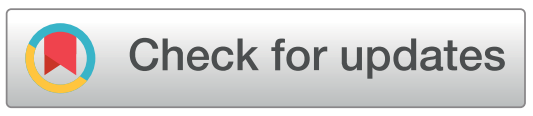

# Preserved metallicity and tunable magnetism in Zr-based Janus MXenes

S. Özcan *[a] and B. Biel [b]

Janus MXenes provide a chemically asymmetric platform to tailor the properties of two-dimensional transition metal carbides. Here, we present a systematic first-principles study of the structural, electronic, mechanical, and magnetic properties of Janus $ZrMCX_2$ MXenes (M = Cr, Hf, Nb, Sc, Ti; X = F, Cl, S). All compositions exhibit negative formation energies and satisfy the Born stability criteria, supporting the energetic and mechanical stability of the Janus structures. In contrast to the metal-to-semiconductor transitions frequently reported for functionalized MXenes, we find that the metallic character is preserved in nearly all $ZrMCX_2$ systems, despite the strong chemical asymmetry introduced by the Janus configuration. Surface functionalization nevertheless plays a key role in modulating the electronic density near the Fermi level and strongly influences the magnetic and mechanical responses. Several compounds exhibit ferromagnetic or antiferromagnetic ground states, three display half-metallicity, and sizable magnetic anisotropy energies are obtained, with Néel temperatures exceeding room temperature in selected cases. Functionalization also enhances the elastic stiffness of the Janus structures. These results demonstrate that chemical asymmetry and surface termination provide effective control over magnetic and mechanical properties while maintaining a robust metallic framework, identifying Zr-based Janus MXenes as promising model systems for metallic and spin-dependent two-dimensional materials.

## 1 Introduction

Two-dimensional (2D) materials have attracted sustained interest due to their unique electronic, mechanical, and chemical properties, which make them promising candidates for applications in energy storage, nanoelectronics, sensing, and catalysis.[1,2] Since the discovery of graphene, a broad range of atomically thin systems—including transition metal dichalcogenides (TMDs),[3] hexagonal boron nitride,[4] silicene,[5] phosphorene,[6] and MXenes[7–9]—has considerably expanded the landscape of low-dimensional materials for potential applications.

MXenes, described by the general formula $M_{n+1}X_nT_x$ (where M is a transition metal, X is carbon or nitrogen, and T denotes surface terminations), are particularly appealing due to their high electrical conductivity, chemical versatility, and mechanical robustness.[1,10,11] A defining feature of MXenes is the possibility of tailoring their properties through surface functionalization, which has been widely reported to induce significant changes in electronic structure, catalytic activity, and ion transport.[11–17,21] In many cases, functionalization with electronegative groups such as O, F, or OH leads to metal-to-semiconductor transitions, especially in Zr- and Ti-based MXenes.[13–15]

Consistent with this behavior, modifying the surface chemical environment is known to have a profound impact on the physical and electronic properties of MXenes. For instance, pristine $Ti_2C$, $V_2C$, $Cr_2C$, $Zr_2C$, $Hf_2C$, and $Ta_2C$ are metallic,[12] whereas several oxygen- or hydroxyl-terminated counterparts, including $Ti_2CO_2$, $Zr_2CO_2$, $Hf_2CO_2$, $Sc_2CO_2$, and $Sc_2C(OH)_2$, exhibit semiconducting behavior.[13] These examples illustrate how surface terminations can drastically alter the electronic character of MXenes.

Beyond electronic structure, surface functionalization also plays a critical role in enabling specific applications. Zhou *et al.*[14] reported the first synthesis of $Zr_3C_2T_2$ (T = –O, –F, –OH), demonstrating their potential as electrode materials for electrical energy storage and as sensing platforms. The surface-functionalized Zr-based MXene $ZrCO_2$ was subsequently investigated theoretically and identified as a promising catalyst material.[15] In the context of energy storage, DFT studies have shown that $Zr_2CX_2$ (X = F, O, S) MXenes exhibit high Li storage capacities and favorable cycling performance. High Na storage capacities were also predicted for $Zr_2CO_2$ and $Zr_3C_2O_2$.[16] Furthermore, the choice of surface termination strongly influences the role of MXenes as electrode

[a] *Department of Physics, Aksaray University, 68100 Aksaray, Turkey. E-mail: sozkaya@aksaray.edu.tr*
[b] *Department of Atomic, Molecular and Nuclear Physics & Instituto Carlos I de Física Teórica y Computacional, Faculty of Science, Campus de Fuente Nueva, University of Granada, 18071 Granada, Spain*

materials, as demonstrated by $Hf_2CT_2$ (T = Cl, OH, F) and pristine $Hf_2C$ as potential anode materials, while $Hf_2CX_2$ (X = O, S, Se) are proposed as cathode materials.[17]

Although *O*-termination is widely reported as the most thermodynamically stable configuration in MXenes,[18] limitations include reduced electrical conductivity, limited experimental control to create purely O-terminated surfaces, and high oxidation rates in humid environments compared to alternative surface terminations. In the present work, we focus on the role of more electronically and chemically active terminations that can tune the material properties, rather than O, such as F, Cl, and S. While the F-terminated MXenes are highly resilient against oxidation,[19] Cl-terminations provide high thermal stability,[20] and *S*-termination, specifically, provides lower diffusion barriers than *O*-terminated counterparts.[21]

Janus 2D materials, characterized by chemically distinct top and bottom surfaces, introduce an additional degree of freedom by breaking out-of-plane mirror symmetry. This asymmetry can modify charge distribution, bonding character, and spin-dependent interactions, leading to novel electronic and magnetic behavior.[1,22–27] Although the experimental realization of Janus MXenes remains limited, recent studies have demonstrated that such structures are feasible. Wet-chemistry etching using F- or Cl-containing aqueous acids produces MXenes with mixed surface terminations (–OH, –O, and –F/–Cl) randomly distributed across their surfaces.[28,29] Another alternative method is the Lewis-acidic melt etching route.[30] Different methods have also been proposed, such as electrochemical and hydrothermal etching by experimentalists.[31,32] Recently, a gas–liquid–solid triphasic etching strategy was developed to produce MXenes with pure and precisely tunable halogen terminations (Cl, Br, I, or their combinations).[33] This method yields MXenes with excellent structural integrity and uniformly ordered surface terminations. Even though fully ordered Janus MXenes have not yet been conclusively demonstrated, the successful synthesis of MXenes with mixed surface chemistries represents an important step towards the experimental realization of Janus structures. Recent theoretical studies on Janus MXenes and related systems have demonstrated enhanced optical responses and tunable catalytic properties, highlighting the potential of chemical asymmetry as a design strategy.[34,35] In addition to optical and catalytic functionalities, Janus engineering has emerged as an effective strategy for tailoring magnetic properties in two-dimensional materials. The broken out-of-plane symmetry can modify exchange interactions, magnetic anisotropy, and spin-dependent electronic states, providing additional degrees of freedom for spintronic applications. Recent studies have demonstrated tunable magnetism and enhanced magnetic anisotropy in Janus and asymmetrically functionalized low-dimensional systems, including MXenes and related compounds.[36–40]

Despite these advances, the combined effects of Janus asymmetry and surface functionalization on the mechanical, electronic, and magnetic properties of Zr-based MXenes remain insufficiently explored. In particular, while functionalization is often discussed in terms of gap opening, much less attention has been paid to situations in which the metallic character is preserved and yet other key properties—such as magnetic order, magnetic anisotropy, and mechanical stiffness—are strongly modified. Understanding this regime is essential for applications that require robust metallic transport together with tunable magnetic functionality.

## 2 Method

All calculations were performed within density functional theory (DFT) using the Vienna *Ab initio* Simulation Package (VASP).[41–44] The exchange–correlation interaction was treated using the Perdew–Burke–Ernzerhof (PBE) functional within the generalized gradient approximation (GGA).[45–47] The Kohn–Sham single-particle wave functions were expanded in a plane-wave basis set with an energy cutoff of 500 eV. Atomic positions were fully relaxed until the Hellmann–Feynman forces on each atom were smaller than 0.001 eV $Å^{-1}$. A Gaussian smearing scheme with a width of 0.05 eV was used to treat partial occupancies.

Brillouin-zone integrations were carried out using a 21 × 21 × 1 Monkhorst–Pack *k*-point grid.[48] To avoid spurious interactions between periodically repeated layers, a vacuum spacing of at least 15 Å was introduced along the direction perpendicular to the MXene plane.

To clarify the role of van der Waals (vdW) interactions, we conducted further test calculations including vdW corrections (DFT-D3).[49] The inclusion of vdW interactions leads to negligible changes in the optimized structural parameters (lattice constants varied by less than 0.02 Å), which is consistent with previous first-principles studies on MXene monolayers, where standard GGA-PBE calculations without vdW corrections provided reliable structural and electronic properties.[15,50] Based on these findings, vdW corrections were omitted.

Electronic band structures were computed using the Heyd–Scuseria–Ernzerhof (HSE06) hybrid functional.[51–53] The HSE06 functional was employed to obtain more accurate band dispersions and band gap values for systems exhibiting semiconducting or half-metallic behavior, while PBE results were used to analyze overall electronic trends.

Formation energies of the functionalized Janus MXenes were calculated to assess their thermodynamic stability according to eqn (1), using the total energies of the pristine ZrMC monolayers, isolated $F_2$ and $Cl_2$ molecules in the gas phase, and bulk sulfur as reference states.

Mechanical properties were evaluated by calculating the elastic constants using the stress–strain method.[54] The mechanical stability of the pristine and functionalized structures was assessed using the Born criteria for hexagonal two-dimensional systems. Young's modulus, in-plane stiffness, shear modulus, and Poisson's ratio were derived from the calculated elastic constants.

Spin-polarized calculations were carried out to determine the magnetic ground states of the Janus $ZrMCX_2$ MXenes. Non-magnetic (NM), ferromagnetic (FM), and three antiferromagnetic configurations (AFM1–AFM3) were considered using a 2 × 2 supercell, which represents the minimum cell size required to capture the relevant magnetic orderings.

Spin–orbit coupling (SOC) was included in noncollinear calculations to evaluate magnetic anisotropy energies (MAEs). The MAE was defined as the energy difference between different spin-orientation directions with respect to the magnetic easy axis.

Magnetic transition temperatures were estimated within a mean-field approximation (MFA) using the energy difference between the magnetic ground state and the lowest-energy competing magnetic configuration (FM or AFM), as described in Section 3.4.

# 3 Results and discussion

## 3.1 Structural properties and formation energies

The optimized crystal structures of the pristine Janus MXenes ZrMC (M = Cr, Hf, Nb, Sc, Ti) are shown in Fig. 1(a) and (b). Their lattice parameters lie in the range 3.17–3.24 Å, and the value obtained for $Zr_2$ C (3.27 Å) agrees well with previous theoretical work.[55] The corresponding Zr–C and M–C bond lengths are listed in Table 1.

The effect of functionalization with F, Cl, and S was examined using three adsorption geometries (fcc, hcp, and fcc + hcp), shown in Fig. 2. In the fcc model, functional groups are located directly below and above the Zr and M atoms, respectively. Meanwhile, in the hcp model, these surface atoms are aligned with the C atoms, and in the fcc + hcp model, one of the functional atoms is placed on the M atom, and the other is below the C atom.

Formation energies, summarized in Table 2, were computed according to eqn (1), where $E_{tot}$ ($ZrMCX_2$), $E_{tot}$ (ZrMC), $E_{tot}$ ($X_2$) are, respectively, the total energy of functionalized ZrMC (f-ZrMC) with termination of $X_2$ (X = F, Cl, S) atoms, the total energy of pristine ZrMC (M = Cr, Hf, Nb, Sc, Ti), and the total energy of $F_2$, $Cl_2$ (in gas phase), and $S_2$ (in stable bulk form). According to the relative energies reported in Table 2, the fcc configuration is the most stable for thirteen systems, while the remaining two adopt the fcc + hcp structure. To assess whether

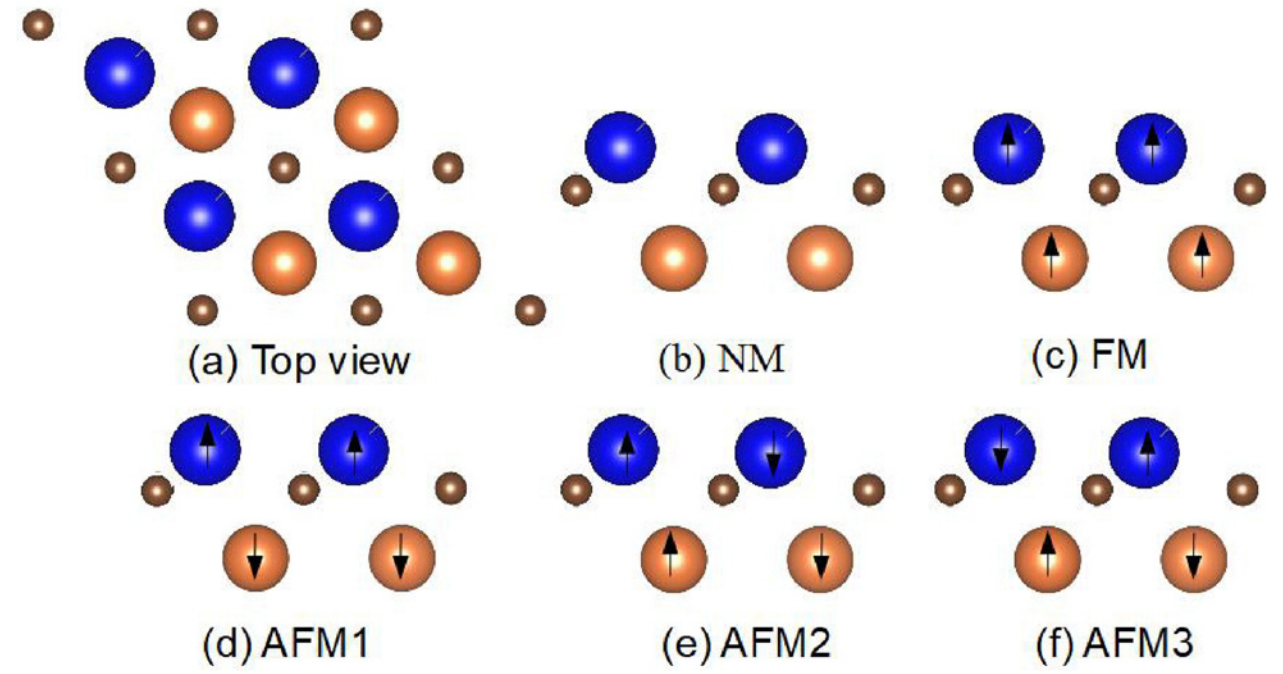

Fig. 1 (a) Top view and side view of ZrMC (M = Cr, Hf, Nb, Sc, Ti) MXene (b) non-magnetic (NM) (c) ferromagnetic (FM), and (d)–(f) antiferromagnetic (AFM) solutions. Zr atoms are shown in blue, M atoms in orange, and C atoms in brown.

Table 1 Calculated lattice parameters, thicknesses $d_{Zr-M,\perp}$, and bond lengths between the Zr (M) and the C atom ($d_{Zr-C}$($d_{M-C}$) of ZrMC (M = Cr, Hf, Nb, Sc, Ti) MXenes

| Material | $a_0$ (Å) | $d_{Zr-M,\perp}$(Å) | $d_{Zr-C}$ (Å) | $d_{M-C}$ (Å) |
|---|---|---|---|---|
| ZrCrC | 3.24 | 2.05 | 2.31 | 2.05 |
| ZrHfC | 3.24 | 2.54 | 2.29 | 2.01 |
| ZrNbC | 3.17 | 2.49 | 2.28 | 2.14 |
| ZrScC | 3.24 | 2.51 | 2.54 | 2.25 |
| ZrTiC | 3.17 | 2.40 | 2.88 | 2.10 |

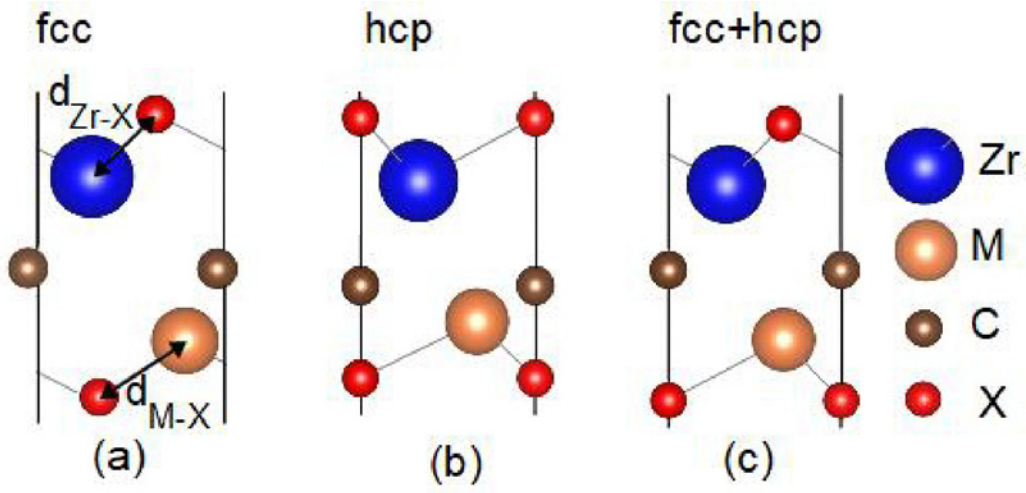

Fig. 2 Side views of (a) fcc, (b) hcp, and (c) fcc + hcp models for $ZrMCX_2$ (M = Cr, Hf, Nb, Sc, Ti; X = F, Cl, S) MXene.

spin polarization affects the relative stability of the candidate adsorption configurations, additional spin-polarized ferromagnetic calculations were performed for the fcc, hcp, and fcc + hcp models of all $ZrCrCX_2$ and $ZrNbCX_2$ (X = F, Cl, S) Janus MXenes. In all cases, the energetic ordering remained unchanged, and the same adsorption geometry was identified as the lowest-energy configuration (Table S2 of the SI). The preservation of the energetic ordering in the spin-polarized calculations indicates that the identification of the preferred adsorption geometries is robust against the inclusion of ferromagnetic spin polarization. In this work, we therefore concentrate on the 15 lowest-energy models, while their magnetic properties are analysed separately in Section 3.4. This section focuses on identifying the preferred structural configuration from total-energy comparisons. In addition, all formation energies are negative, confirming high thermodynamic stability for every composition considered. The magnitude of the formation energies becomes more negative following the electronegativity trend F > Cl > S. Consistently, Zr–X and M–X bond lengths reflect stronger bonding for F-functionalized systems, as shown in Table 2. These negative formation energies confirm that functionalization provides a stabilizing effect on all Janus ZrMC structures.

$$\Delta E_f = E_{tot}(ZrMCX_2) - E_{tot}(ZrMC) - E_{tot}(X_2) \quad (1)$$

## 3.2 Electronic properties

Band structure calculations using the HSE06 functionals show that all pristine ZrMC MXenes are metallic, as shown in SI (Fig. S1). A remarkable outcome of our analysis is that this metallic character is largely preserved upon Janus functionalization: 14 out of the 15 $ZrMCX_2$ compounds remain metallic, despite the strong chemical asymmetry introduced by the Janus configuration and the presence of electronegative surface groups. This behavior contrasts with several previous studies on functionalized MXenes—including Zr-based systems—in which O-, F-, or OH-termination frequently induces semiconducting states or significantly alters the band dispersion.[11–17,21,35,56,57] The robustness of the metallic character observed here suggests that the Zr–C and M–C frameworks in these Janus structures remain electronically dominant even under pronounced surface perturbations.

Table 2 Lattice parameters $a$ (in Å), bond lengths ($d_{Zr-X}$, $d_{M-X}$), thickness of the MXene layers ($t$), relative energies with respect to the minimum energy for each model $\Delta E$ (in eV), formation energies $E_f$ (in eV), and band gaps (in eV) for the $ZrMCX_2$ (M = Cr, Hf, Nb, Sc, Ti, X = F, Cl, S) in the four possible models. Here, M and D indicate the metallic or direct bandgap semiconductor character, respectively. $E_f$ and $E_g$ are shown only for the most stable model of each structure

| | FCC | | HCP | | FCC + HCP | | | | | | |
|---|---|---|---|---|---|---|---|---|---|---|---|
| Material | $a$ | $\Delta E$ | $a$ | $\Delta E$ | $a$ | $\Delta E$ | $d_{Zr-X}$ | $d_{M-X}$ | $t$ | $E_f$ | $E_g$ |
| $ZrCrCF_2$ | 3.23 | 0.00 | 3.23 | 1.32 | 3.23 | 0.78 | 2.30 | 2.22 | 4.62 | −9.96 | 1.68 (D) |
| $ZrCrCCl_2$ | 3.23 | 0.00 | 3.23 | 0.83 | 3.23 | 0.54 | 2.63 | 2.48 | 5.58 | −5.87 | M |
| $ZrCrCS_2$ | 3.23 | 0.27 | 3.23 | 0.47 | 3.23 | 0.00 | 2.51 | 2.35 | 5.51 | −4.77 | M |
| $ZrHfCF_2$ | 3.23 | 0.00 | 3.23 | 0.24 | 3.23 | 0.06 | 2.32 | 2.32 | 5.19 | −11.44 | M |
| $ZrHfCCl_2$ | 3.23 | 0.01 | 3.23 | 0.05 | 3.23 | 0.00 | 2.64 | 2.63 | 6.23 | −6.92 | M |
| $ZrHfCS_2$ | 3.23 | 0.00 | 3.23 | 0.89 | 3.23 | 0.47 | 2.53 | 2.51 | 6.08 | −5.27 | M |
| $ZrNbCF_2$ | 3.18 | 0.00 | 3.18 | 0.47 | 3.18 | 0.31 | 2.29 | 2.31 | 5.23 | −10.73 | M |
| $ZrNbCCl_2$ | 3.18 | 0.00 | 3.18 | 0.19 | 3.18 | 0.14 | 2.63 | 2.61 | 6.21 | −6.26 | M |
| $ZrNbCS_2$ | 3.18 | 0.00 | 3.18 | 0.33 | 3.18 | 0.08 | 2.52 | 2.46 | 6.00 | −4.95 | M |
| $ZrScCF_2$ | 3.23 | 0.00 | 3.23 | 0.59 | 3.23 | 0.49 | 2.30 | 2.20 | 5.04 | −11.93 | M |
| $ZrScCCl_2$ | 3.23 | 0.00 | 3.23 | 0.20 | 3.23 | 0.18 | 2.63 | 2.54 | 6.12 | −6.92 | M |
| $ZrScCS_2$ | 3.23 | 0.00 | 3.23 | 0.79 | 3.23 | 0.3 | 2.57 | 2.50 | 6.13 | −4.30 | M |
| $ZrTiCF_2$ | 3.18 | 0.00 | 3.18 | 0.43 | 3.18 | 0.23 | 2.29 | 2.20 | 4.98 | −11.31 | M |
| $ZrTiCCl_2$ | 3.18 | 0.00 | 3.18 | 0.16 | 3.18 | 0.11 | 2.63 | 2.50 | 5.99 | −6.68 | M |
| $ZrTiCS_2$ | 3.18 | 0.00 | 0.86 | 3.18 | 0.42 | 3.18 | 2.54 | 2.40 | 5.90 | −4.87 | M |

Only one system, $ZrCrCF_2$, deviates from this general trend and becomes a direct semiconductor with an HSE06 band gap of 1.68 eV, as shown in the SI (Fig. S1) and summarized in Table 2. All other functionalized systems retain metallic features across the Brillouin zone.

The projected density of states (Fig. 3) further confirms that the states near the Fermi level are dominated by hybridized Zr-d and M-d orbitals. For Cr-containing MXenes, Cr-d contributions remain significant in both valence and conduction bands, in agreement with their magnetic behavior discussed in Section 3.4.

To further analyze the orbital hybridization between constituent atoms, we examined the electron localization functions (ELF)[58] within the (110) plane, and renormalized the value to 0.00–0.9 for the $ZrMCX_2$ MXenes. A value of 0.0 corresponds to a very low charge density, while 0.9 and 0.45 correspond to fully localized (covalent-bonding) and delocalized (metallic-bonding) electronic distributions, respectively. The ELF maps (Fig. S2) show a cloud of lone-pair electrons, mainly located around the terminal atoms (X = F, Cl, S). The relatively delocalized ELF distribution within the transition-metal network is consistent with the metallic character revealed by the band-structure and density-of-states calculations, while localized charge accumulation is mainly observed around the terminal atoms.

It is worth noting that the preservation of metallicity across most $ZrMCX_2$ compositions does not diminish the role of functionalization. Although the electronic character remains metallic, the surface groups significantly modify the electronic density near the Fermi level, alter the hybridization between Zr-d and M-d states, and enable the emergence or modulation of magnetic ordering. Functionalization also enhances the mechanical robustness of the Janus structures, as reflected in the increased elastic constants, and introduces chemical asymmetry that may be relevant for adsorption or catalytic processes. Thus, the functional groups act as effective tuning parameters that adjust electronic, magnetic, and mechanical properties while preserving the desirable metallic framework of the pristine MXenes.

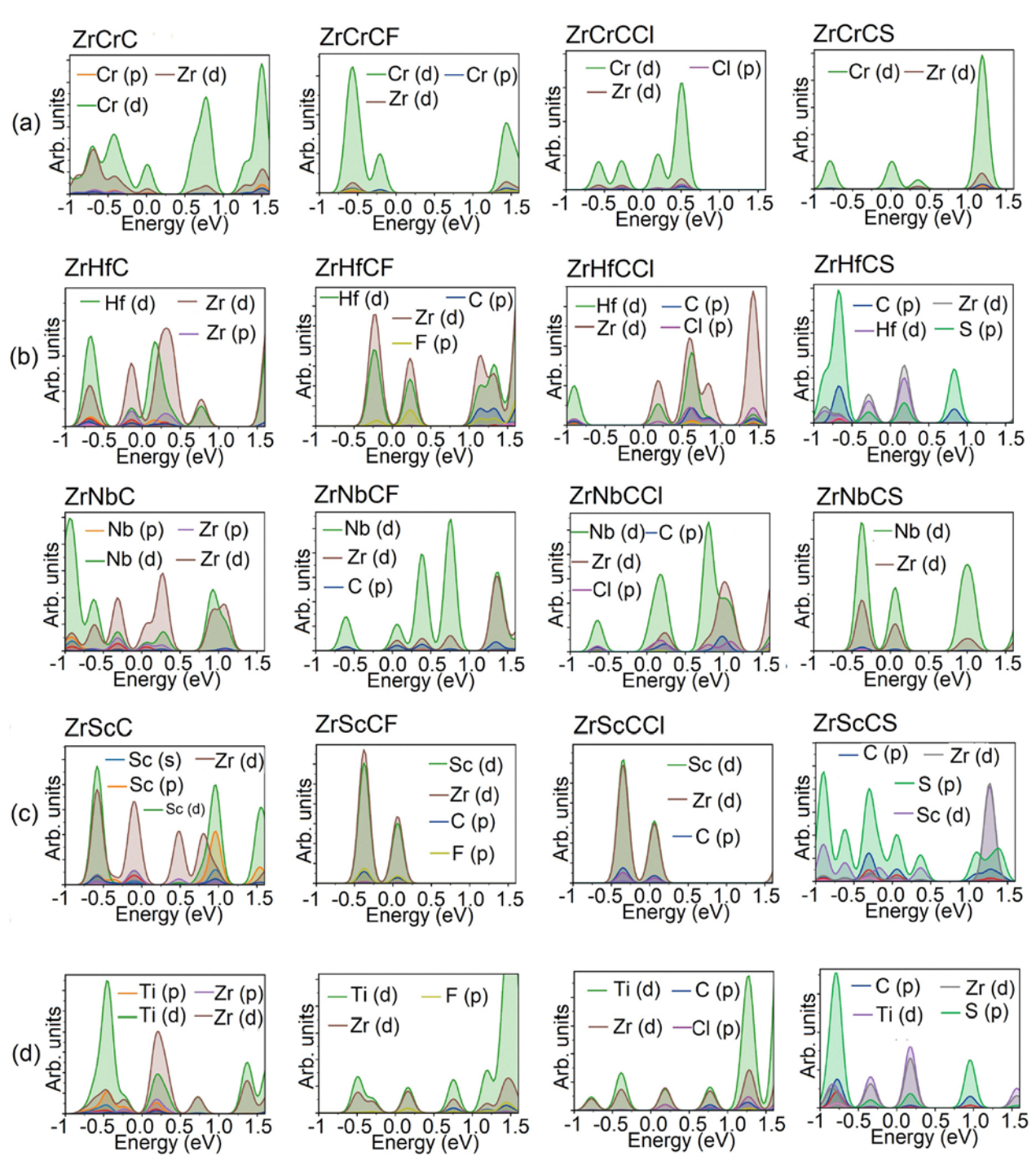

Fig. 3 DOS structures of (a) $ZrCrCX_2$, (b) $ZrHfCX_2$, (c) $ZrNbCX_2$, (d) $ZrScCX_2$, (e) $ZrTiCX_2$ (X = F, Cl, S). The Fermi energy is set at zero.

### 3.3 Mechanical stability and elastic properties

Mechanical stability was assessed using the Born criteria[59] following eqn (2), and all pristine and functionalized structures satisfy the required conditions.

The elastic constants are summarized in Table S1. In hexagonal symmetry, $C_{11}$ and $C_{12}$ ($C_{11} > C_{12}$) are the two independent nonzero second-order elastic constants, with $C_{11}$ describing the response to uniaxial stress along the $\langle 100 \rangle$ direction. As shown in Table S1., $C_{11}$ ranges from 135 to 310 N m$^{-1}$, and functionalization generally increases both $C_{11}$ and $C_{12}$. The highest stiffness is obtained for $ZrNbCS_2$ ($C_{11}$ = 309.59 N m$^{-1}$), while ZrCrC is the least stiff. Overall, the stiffness of the MXenes studied here is lower than that of graphene (341 N m$^{-1}$)[60] but exceeds that of typical TMDs (*e.g.*, $WS_2$ : 139 N m$^{-1}$).[60] Because $C_{12}$ describes the deformation response in the (110) plane, changes in this parameter reflect the influence of surface chemistry on mechanical behavior. As observed in Table S1., functionalization leads to a notable increase in $C_{12}$ across Zr-based Janus MXenes; for instance, ZrHfC exhibits an enhancement from 63.806 N m$^{-1}$ to 112.384 N m$^{-1}$ when S atoms are introduced.

Young's modulus ($Y$), in-plane stiffness ($B$), shear modulus ($G$), and Poisson's ratio ($\nu$) can be calculated from the elastic constants ($C_{ij}$) by means of eqn (3). We note that all 15 Janus $ZrMCX_2$ MXenes exhibit an increase in Young's modulus, ranging from 2% to 79%, compared to their pristine ZrMC counterparts (see Table S1). These results clearly demonstrate that functionalization consistently enhances the elastic stiffness of the materials. Computed Young's moduli range from 116 to 270 N m$^{-1}$, exceeding those of several 2D materials such as $MoSe_2$ (103.9 N m$^{-1}$),[60] $MoTe_2$ (79.4 N m$^{-1}$),[60] $WTe_2$ (86.4 N m$^{-1}$),[60] silicene (62 N m$^{-1}$),[61] while remaining below that of graphene (341 N m$^{-1}$).[60] Poisson's ratio $\nu$, usually ranges between −1 and 0.7, and it shows low values for pure covalent materials, such as graphene ($\nu$ = 0.169[62]). The higher value of $\nu$ indicates a greater tendency for lateral deformation under applied strain. In our study, Poisson ratios vary between 0.3 and 0.53, with the highest value (0.532) for $ZrCrCF_2$, indicating ductile mechanical behavior. Functionalization also enhances the in-plane stiffness $B$, *e.g.* increasing from 127 N m$^{-1}$ in ZrNbC to 210 N m$^{-1}$ in $ZrNbCS_2$.

These trends indicate that surface functionalization not only preserves the mechanical stability of the Janus ZrMC frameworks—as all systems satisfy the Born criteria—but also enhances their elastic response, as reflected in the systematic increase of $C_{11}$, $C_{12}$, $Y$, and $B$ without introducing any mechanical anomalies. Taken together, the negative formation energies, satisfaction of the two-dimensional elastic-stability criteria, and robustness of the structural ordering under spin-polarized calculations support the energetic and mechanical viability of the proposed Janus structures. Nevertheless, these results do not constitute a complete assessment of dynamical or finite-temperature stability. Such an assessment would require systematically converged phonon-dispersion and/or molecular-dynamics calculations, which are beyond the scope of the present screening study and should be addressed in future work.

$$C_{11} > 0,\ C_{66} > 0, 2{*}C_{66} = C_{11} - C_{12},\ \text{and}\ C_{11} > |C_{12}| \quad (2)$$

$$Y = (C_{11}^2 - C_{12}^2)/C_{11},\ \nu = C_{12}/C_{11},\ G = (C_{11} - C_{12})/2,\ B = Y/2{*}(1 - \nu) \quad (3)$$

### 3.4 Magnetic properties

To determine the magnetic behavior of the Janus $ZrMCX_2$ MXenes, we compared non-magnetic (NM), ferromagnetic (FM), and three antiferromagnetic (AFM1–AFM3) configurations using a 2 × 2 supercell (Fig. 1(c)–(f)). Eight systems exhibit magnetic ground states: ferrimagnetic ZrCrC, ferromagnetic $ZrCrCF_2$, $ZrCrCCl_2$ and ZrNbC, antiferromagnetic $ZrCrCS_2$, ZrHfC, ZrScC and ZrTiC, while the remaining compounds are non-magnetic (see Table S3). The Cr-based MXenes show the richest magnetic behavior, with their magnetic configuration strongly influenced by the surface functional group.

$ZrCrCF_2$ and $ZrCrCCl_2$ stabilize in FM ground states with a total magnetic moment of $8\mu_B$ per supercell, and ZrNbC is also FM with a moment of $2.4\mu_B$. In contrast, ZrHfC, ZrScC, and ZrTiC favor the AFM1 ordering, presenting supercell moments of 1.73, 1.39, and $-1.93\mu_B$, respectively. $ZrCrCS_2$ adopts an AFM3 ground state with zero net magnetization. The local magnetic moments on the Zr atoms ($0.262\mu_B$) are significantly smaller than those on the Cr atoms ($2.712\mu_B$) and are aligned antiparallel to them. Although the lowest-energy spin arrangement corresponds to the AFM1 pattern, the unequal compensation between the Cr and Zr magnetic moments results in a large net magnetic moment of $-9.39\mu_B$ per supercell. Therefore, ZrCrC should be classified as a ferrimagnetic (FiM) system rather than a conventional antiferromagnet. A similar FiM character arising from antiferromagnetically coupled magnetic sublattices has been reported in previous theoretical studies of CrScC[63] and $Cr_2VC_2F_2$.[64]

These trends highlight that functionalization, although it preserves metallicity in most systems, plays a decisive role in stabilizing specific magnetic arrangements by modifying the electronic density near the Fermi level and reshaping the Zr-d/M-d hybridization.

Spin-polarized band structures (Fig. 4) show that several magnetic MXenes, such as ZrCrC, ZrNbC, ZrScC, and ZrTiC, remain metallic in both spin channels. However, three compounds display half-metallicity: $ZrCrCF_2$ (spin-up gap of 0.83 eV), $ZrCrCCl_2$ (spin-up gap of 0.55 eV), and ZrHfC (spin-down gap of 0.18 eV). The substantial half-metallic gaps in $ZrCrCF_2$ and $ZrCrCCl_2$ indicate a strong spin polarization at the Fermi level and suggest robustness of the half-metallic character against thermal broadening effects. These results confirm that surface functionalization enables fine control over spin transport channels without disrupting the metallic backbone of the MXenes.

To examine the robustness of half-metallicity to structural perturbations, we performed test calculations under biaxial strain. ZrHfC retains its half-metallic character under biaxial strains ranging from −3% to +3%. While slight variations in the half-metallic gap are observed (see Fig. S3), no transition to a metallic or semiconducting state occurs within this range, indicating that half-metallicity remains largely preserved under moderate strain. In the case of $ZrCrCF_2$ and $ZrCrCCl_2$, under biaxial strain, the energy difference between the FM and AFM1 states decreases to a narrow range of a few meV, while the half-metallic electronic structure remains intact. Previous studies have shown that external strain can significantly vary exchange interactions, leading to AFM–FM transitions or strong magnetic competition in various MXenes such as $Zr_2$ N, $Ti_2C$, and related Janus systems.[65–67] These results show that although half-metallicity remains robust, the magnetic ground state may become sensitive to external perturbations, which could be advantageous for tunable spintronics applications.

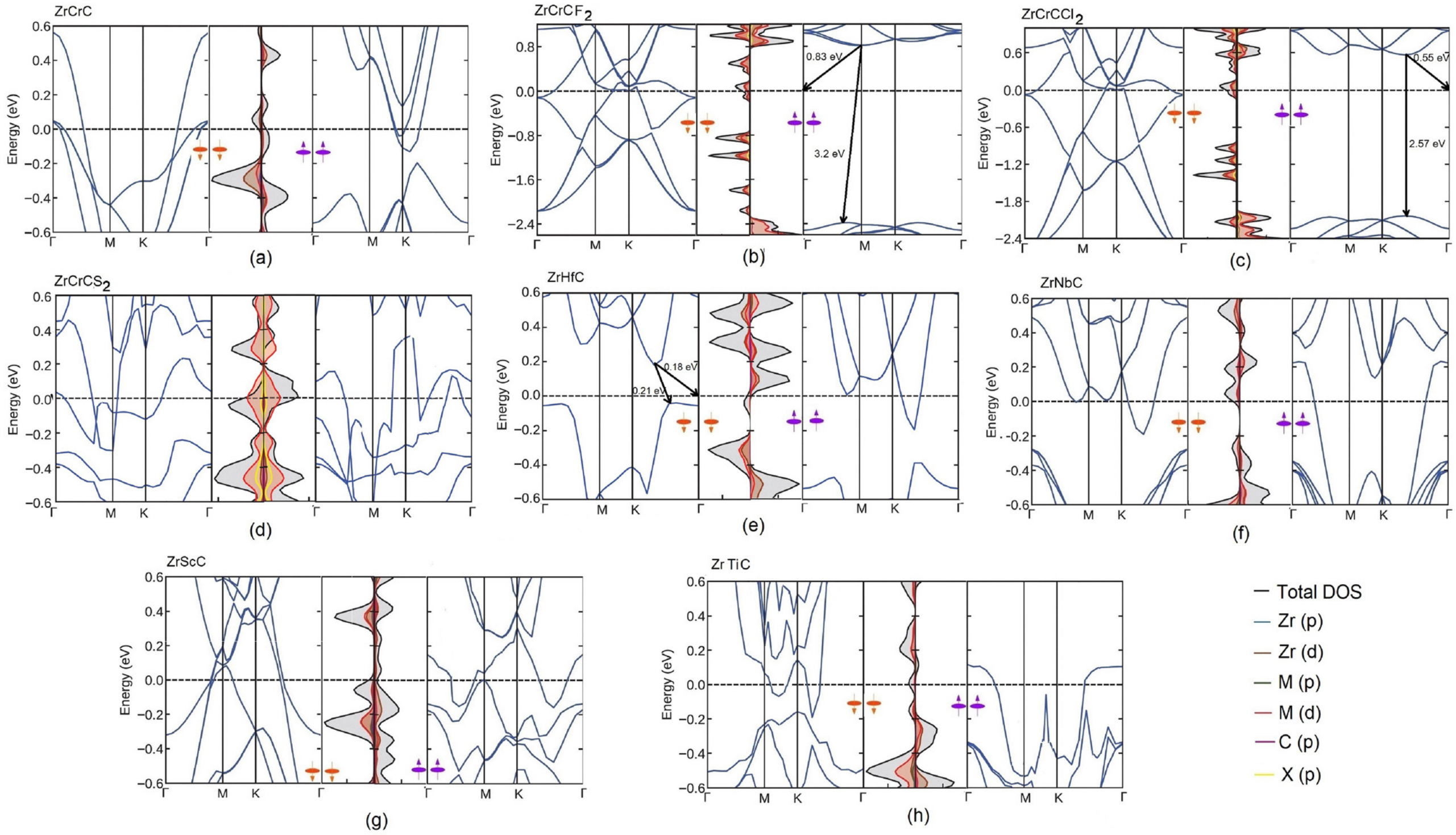

Fig. 4 The band and spin-resolved density of states structures of (a) ZrCrC, (b) $ZrCrCF_2$, (c) $ZrCrCCl_2$, (d) $ZrCrCS_2$, (e) ZrHfC, (f) ZrNbC, (g) ZrScC, and (h) ZrTiC. The Fermi level is set at 0 eV. Bands and DOS for spin-down (up) are shown on the left (right) panels. The black arrow shows the HSE06 band gap.

Magnetic anisotropy energies (MAE) were computed to determine the preferred spin orientation. $ZrCrCF_2$, $ZrCrCCl_2$, and $ZrCrCS_2$ exhibit in-plane easy axes, whereas ZrCrC, ZrHfC, ZrScC, ZrNbC, and ZrTiC favor out-of-plane spin alignment. The magnitude of the MAEs is significant, reaching 469 μeV for $ZrCrCF_2$, 208 μeV for ZrCrC, and 129 μeV for ZrScC. Such values exceed those of many layered magnetic materials and point to strong resistance against thermal spin fluctuations.

Magnetic transition temperatures were estimated using the mean-field approximation (MFA), $k_BT_C = (2/3)\Delta(E)$.[68] Here, $k_B$ and $\Delta(E)$ are the Boltzmann constant and the energy difference between the FM and the most stable AFM configuration, respectively. It should be noted that the mean-field approximation neglects thermal spin fluctuations and therefore generally overestimates magnetic transition temperatures in low-dimensional systems. Consequently, the values reported here should be regarded as upper-bound estimates rather than quantitative predictions. $ZrCrCF_2$, $ZrCrCCl_2$, ZrNbC, and ZrScC exhibit relatively low Curie or Néel temperatures in the range of 8–15 K. These values reflect the limitations of a mean-field description for low-dimensional magnetic systems, particularly in cases where competing magnetic configurations are close in energy. Importantly, however, the sizable energy differences between the non-magnetic and magnetic solutions indicate that the formation of local magnetic moments is robust.

ZrHfC exhibits a substantially higher Néel temperature, while ZrTiC and ferrimagnetic ZrCrC display intermediate ordering temperatures. These estimated values exceed those reported for several well-known two-dimensional magnetic materials, *e.g.*, $CrI_3$ (45 K),[69] $HfI_3$ (78 K),[69] $Sc_2COF$ (110.5 K),[34] and point to enhanced magnetic stability driven by the combined effects of Janus asymmetry and transition-metal chemistry. For clarity, a summary of the main structural/electronic trends and the magnetic descriptors is provided in Tables 3 and 4.

Table 3 Summary of structural, electronic, and magnetic properties of calculated structures in this study

| Property | F | Cl | S | Strongest effect |
|---|---|---|---|---|
| Stability ($E_f$) | ↑↑ | ↑ | ↓ | F strongest |
| Metal → semiconductor | Only $ZrCrCF_2$ | None | None | F strongest |
| Elastic constant $C_{11}$ | ↑ | ↑↑ | ↑ | Cl strongest in many cases |
| Magnetic behavior | FM/AFM changes | Varies | Varies | Cr-based systems |

Table 4 Magnetic properties overview

| MXene | Ground state | Conductivity | $T_{C/N}$ | MAE direction |
|---|---|---|---|---|
| ZrCrC | FiM | Metal | 212 | 001 |
| $ZrCrCF_2$ | FM | Half-Metal | 8 | 100 |
| $ZrCrCCl_2$ | FM | Half-Metal | 8 | 100 |
| $ZrCrCS_2$ | AFM3 | Metal | 15 | 100 |
| ZrHfC | AFM1 | Half-Metal | 500 | 001 |
| ZrNbC | FM | Metal | 8 | 001 |
| ZrScC | AFM1 | Metal | 8 | 001 |
| ZrTiC | AFM1 | Metal | 185 | 001 |

Overall, functionalization and Janus asymmetry provide a versatile mechanism for tuning magnetic order, half-metallicity, and anisotropy in Zr-based MXenes. The coexistence of robust metallicity, tunable magnetic ground states, large MAEs, and high transition temperatures in selected compounds underscores the potential of these materials for spintronic and magnetic-device applications.

## 4 Conclusions

In this work, we have carried out a comprehensive first-principles investigation of the structural, electronic, mechanical, and magnetic properties of Janus $ZrMCX_2$ MXenes, where M = Cr, Hf, Nb, Sc, Ti and X = F, Cl, S. All studied compositions exhibit negative formation energies and satisfy the Born stability criteria, supporting the energetic and mechanical viability of the proposed structures within the level of theory employed. Moreover, functionalization generally enhances the elastic response, increasing the elastic constants and in-plane stiffness without compromising mechanical stability.

A central finding of this study is that, in contrast to the metal-to-semiconductor transitions frequently reported for functionalized MXenes, the metallic character is preserved in nearly all $ZrMCX_2$ systems, despite the strong chemical asymmetry associated with the Janus configuration. While the metallic framework remains robust, surface functionalization substantially modifies the electronic density near the Fermi level and the hybridization of Zr-d and M-d states, providing an effective means to tune electronic features without disrupting metallic transport.

The Janus Zr-based MXenes also exhibit a rich diversity of magnetic behavior. Depending on the transition metal and surface termination, ferromagnetic or antiferromagnetic ground states are stabilized, and several systems display half-metallicity with sizable spin-dependent gaps. Significant magnetic anisotropy energies are obtained, and enhanced magnetic stability is predicted for selected compounds, highlighting the strong coupling between chemical asymmetry, electronic redistribution, and magnetic ordering in these materials.

Taken together, our results demonstrate that Janus asymmetry and surface functionalization provide an effective strategy to modulate magnetic and mechanical properties while preserving a robust metallic backbone in Zr-based MXenes. This combination of structural stability, metallicity, and tunable magnetism establishes Janus $ZrMCX_2$ MXenes as a versatile platform for exploring structure–property relationships and spin-dependent phenomena in metallic two-dimensional materials. A complete assessment of dynamical and finite-temperature stability will nevertheless require systematically converged phonon or molecular-dynamics calculations.

## Author contributions

Sibel Özcan: conceptualization (lead); formal analysis (lead); writing – original draft (lead); writing – review and editing (equal). Blanca Biel: conceptualization (supporting); writing – review and editing (equal).

## Conflicts of interest

There are no conflicts to declare.

## Data availability

The data supporting the findings of this study are available within the article and its supplementary information (SI). Supplementary information is available. See DOI: https://doi.org/10.1039/d6cp00440g.

## Acknowledgements

The Albaicín supercomputer of the University of Granada and TUBITAK ULAKBIM, High Performance and Grid Computing Center (TRUBA resources) are also acknowledged for providing computational time and facilities. The authors acknowledge TÜBİTAK ULAKBİM EKUAL for supporting the open access publication of this article.